\documentclass[twocolumn,showpacs,preprintnumbers,amsmath,amssymb]{revtex4-1}

\usepackage{graphicx}
\usepackage{amssymb}
\usepackage{hyperref}

\begin{document}

\title{The role of the entropy in the ground state formation of
magnetically frustrated systems within their quantum critical
regime}

\author{Julian G. Sereni}
\address{Low Temperature Division, CAB-CNEA and Conicet, 8400 S.C. de Bariloche, Argentina}

\begin{abstract}

A systematic modification of the entropy trajectory ($S_m(T)$) is
observed at very low temperature in magnetically frustrated
systems as a consequence of the constraint ($S_m\geq 0$) imposed
by the third law of thermodynamics. The lack of magnetic order
allows to explore some unusual thermodynamic properties by tracing
the physical behavior of real systems. The most relevant findings
are: i) a common $C_m/T|_{T\to 0} \approx 7$\,J/molK$^2$ 'plateau'
in at least five Yb-based very-heavy-fermions (VHF) compounds; ii)
quantitative and qualitative differences between VHF and standard
non-Fermi-liquids. iii) Entropy-bottlenecks governing the change
of $S_m(T)$ trajectories in a continuous transition into
alternative ground states that exhibits third order
characteristics. An empirical analysis of the possible $S_m(T\to
0)$ dependencies according to the $\partial ^2 S_m/\partial T^2$
derivative is also preformed. Altogether, this work can be
regarded as an empirical application of the third law of
thermodynamics.

\end{abstract}

\date{\today}

\maketitle

%E-mail: jsereni@cab.cnea.gov.ar

\section{Introduction}

The intensive study of the quantum critical (QC) behavior in heavy
fermion (HF) compounds \cite{Stewart01,QPT07} has powered the
investigation of thermal properties in a significant number of new
Ce and Yb intermetallics at $T\leq 1$\,K. Despite of the
unattainability of the $T=0$ limit, the extremely low
characteristic energies of these systems provide a fertile field
to allow to perform an original test for the applicability of the
third law of thermodynamics in real systems. As a consequence of
this newly available information unpredicted behaviors of the
thermal dependence of the entropy ($S_m$) have emerged in the
thermal range where thermodynamic and QC fluctuations strongly
interplay to stabilize the ground states of not ordered systems.
These topics became recently relevant in the search for new
materials suitable for adiabatic demagnetization refrigeration at
the milikelvin range of temperature \cite{White}

\begin{figure}[tb]
\begin{center}
\includegraphics[width=20pc]{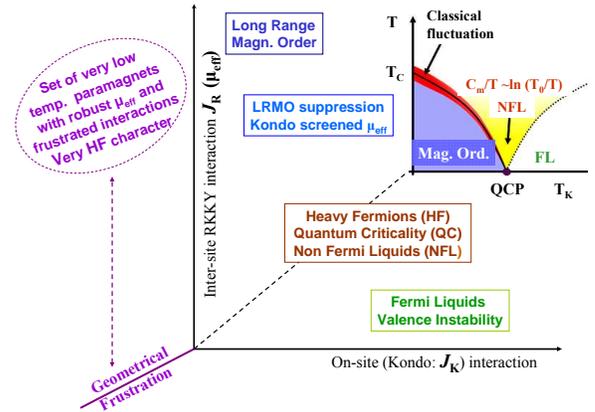}
\end{center}
\caption{(Color online) Schematic description of the magnetic
behavior of Ce- (Yb-) lattice compounds as a function of two
exchange parameters (see the text). The third axis represents
frustrated (paramagnetic) systems. Inset: usual representation as
a function of Kondo temperature. \cite{Doniach,Lavagna}.}
\label{F1}
\end{figure}

The usual magnetic behavior of Ce and Yb compounds can be properly
described as a function of two coupling parameters
\cite{Doniach,Lavagna}: the inter-site RKKY magnetic interaction
($J_{R}$) and the on-site Kondo ($J_K$) interaction. As the local
$J_K$ coupling between band and localized $4f$ states increases,
the intensity of the magnetic moments ($\mu_{eff}$) decreases
because of the Kondo screening. The consequent weakening of
$J_{R}$ ($\propto \mu_{eff}$) is reflected in the decrease of the
magnetic order temperature ($T_{ord}$) that can be driven down to
a QC point \cite{TVojta}. The different stages of this
demagnetizing process as a function of $J_K$ is schematically
resumed in Fig.~\ref{F1}.

Once $T_{ord}$ reaches the range at which thermal fluctuations
compete with quantum fluctuations the QC scenario sets on (see the
inset in Fig.~\ref{F1}). This regime is observed in heavy fermion
(HF) compounds which behave as non-fermi-liquids (NFL) while
approaching the QC regime \cite{Stewart01,QPT07}. Once $J_K$
overcomes $J_R$ the Fermi Liquid (FL) behavior takes over. Within
this regime the thermal ($\gamma = C_m/T$), magnetic ($\chi_0$)
and transport ($\rho = A T^2$) proportionality: $\gamma \propto
\chi_0 \propto \surd A \propto m_{eff}$ is fulfilled, being
$m_{eff}$ the enhanced effective electron mass.

There is, however, an increasing set of Ce and Yb compounds which
escape for this description because they do not order magnetically
despite of their robust $\mu_{eff}$ (i.e. $J_R \gg J_K$) in a
lattice arrangement. This characteristic is mainly due to the
frustration of antiferromagnetic (AF) $J_R$ between magnetic
neighbors favored by some types of atomic coordinations. These
peculiar group is included in the expanded phase diagram presented
in Fig.~\ref{F1} introducing a third axis.

\begin{figure}[tb]
\begin{center}
\includegraphics[width=20pc]{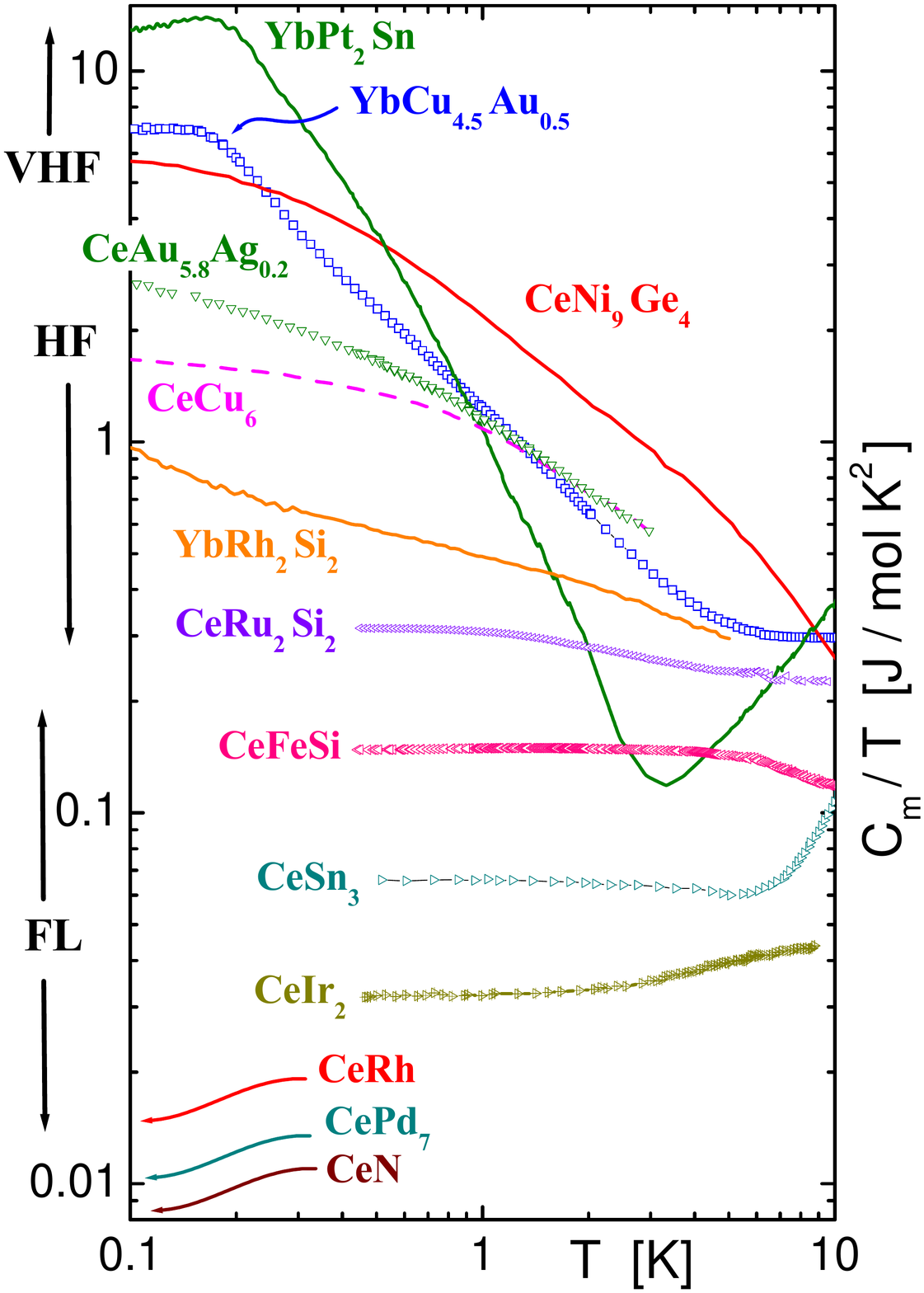}
\end{center}
\caption{(Color online) Examples of measured $C_m/T|_{T\to 0}$
values within four decades, identifying three groups according
their different behaviors: fermi liquids (FL), heavy fermions (HF)
and very heavy fermions (VHF), see the text. For binary FL, see
references in \cite{Handb} and for ternary compounds see
\cite{YbPt2Sn,YbCu5-xAux,CeNi9Ge4,CeCu6}.} \label{F2}
\end{figure}

\section{Different classes of heavy fermions}

Among the Ce and Yb compounds which do not order magnetically, the
measured values of specific heat at $T\to 0$ ($C_m/T|_{T\to 0}$)
cover more than three decades between the lowest reported value
$\approx 8$\,mJ/molK$^2$ in CeN and CePd$_7$ \cite{Handb} and the
highest $\approx 12$\,J/molK$^2$ observed in YbPt$_2$Sn
\cite{YbPt2Sn}, see Fig.~\ref{F2}. CeN and CePd$_7$ are
characterized by a strong local ($4f$) and conduction states
hybridization ($\propto J_K$) reflected in a large Kondo
temperature [$T_K \propto \exp(-1/J_K)$]. On the weak
hybridization range, the exemplary FL-HF with the highest
$C_m/T|_{T\to 0}$ is CeCu$_6$ with a $\gamma=1.6$\,J/molK$^2$
\cite{CeCu6} and $T_K\approx 10$\,K. Its FL character is proved by
the low temperature $\rho = AT^2$ dependence.

Between $1\leq C_m/T|_{T\to 0} \leq 3$\,J/molK$^2$ the most
frequent behavior corresponds to compounds showing the
characteristic NFL dependence: $C_m /T \propto -\ln (T/T_0)$
\cite{Stewart01}, where the energy scale $T_0$ plays a similar
role than $T_K$. Coincidentally, the most representative NFL
compounds show a $\rho\propto T$ dependence \cite{Stewart}.

Recently, some Ce and a significant number of Yb compounds were
found to clearly exceed the $C_m/T|_{T\to 0}$ values of NFL. These
compounds can be labelled as Very-HF (VHF) because their
$C_m/T|_{T\to 0}$ range between $\approx 5$ and $\approx
12$\,J/molK$^2$, see Fig.~\ref{F2}. The absence of magnetic order
in these systems coincides with a huge increase of their spin
correlations (sc) in the paramagnetic (pm) phase by decreasing
temperature. Since at that range of temperature their magnetic
behavior does not fit into the typical Curie-Weiss paramagnetism,
they will be hereafter identified as 'sc-pm'. The strong increase
of the density of magnetic excitations is reflected in a divergent
power law dependence of $C_m(T)/T\propto 1/T^Q$, with exponents
ranging between $1\leq Q\leq 2$. This behavior is observed down to
a temperature ($T_{BN}$) below which a clear deviation from the
sc-pm behavior occurs due to an entropy bottleneck (BN) formation,
to be described below.

\begin{figure}[tb]
\begin{center}
\includegraphics[width=19pc]{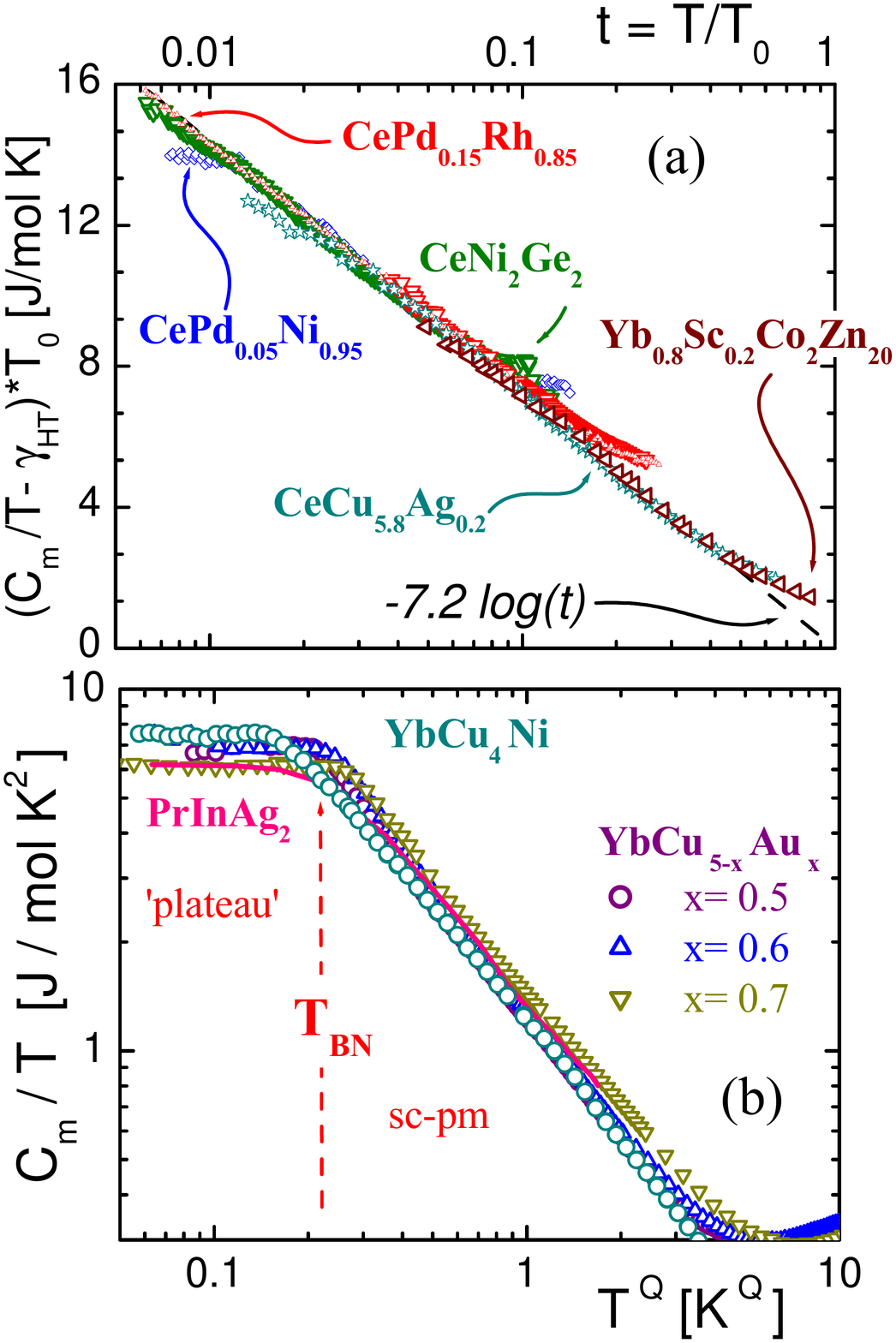}
\end{center}
\caption{(Color online) (a) Overlap of $C_m(t)$ dependencies of
NFL-HF as a function of $-\ln(t)$ using a normalized temperature
$t=T/T_0$ \cite{1997}, where $\gamma_{HT}$ accounts for the high
temperature $C_m/T$ contribution. (b) Power law overlap of five
VHF as a function of normalized $Q$ exponent of $1/T^Q$ in a
double logarithmic representation. The 'sc-pm' region identifies
the spin correlated-paramagnetic phase and $T_{BN}$ the
temperature of the onset of the $C_m(T)/T$ 'plateau' regime. }
\label{F3}
\end{figure}

Apart from the different values of $C_m/T|_{T\to 0}$ between
NFL-HF and VHF, there are other distinctive properties indicating
that these materials belong to different classes. An intrinsic
difference is that in NFL the magnetic moments are increasingly
quenched by Kondo effect and therefore located into the QC region,
see the schematic phase diagram in the inset of Fig.~\ref{F1}. On
the contrary, VHF exhibit robust moments with irrelevant Kondo
effect, being the geometrical frustration responsible for the lack
of magnetic order. As a consequence, their respective thermal
scalings are different. While the $C_m(T)/T\propto -\ln (T/T_0)$
dependence of NFL was found to scale among different compounds
through their $T_0$ temperatures \cite{1997}, see Fig.~\ref{F3}a,
the scaling among VHF compounds can be done through the exponents
$Q$ of their $1/T^Q$ dependencies. This feature is presented in
Fig.~\ref{F3}b for five Yb-based examples using the following
values: $Q = 1\pm 0.2$ for YbCu$_{5-x}$Au$_x$ ($0.5\leq x \leq
0.7$) \cite{YbCu5-xAux}, $Q = 1.4$ for PrInAg$_2$ \cite{PrInAg2},
and $Q = 1.2$ for the YbCu$_4$Ni \cite{Strydom} compound. Notably,
all these compounds show a nearly coincident $C_m/T|_{T\to 0}
\approx 7\pm 0.7$\,J/mol \,K$^2$ 'plateau' below a characteristic
temperature $T_{BN}$.

There are other two compounds belonging to this VHF group: YbBiPt
\cite{YbBiPt} and YbCo$_2$Zn$_{20}$ \cite{YbCo2Zn20}. Although
they show the same $C_m/T|_{T\to 0}$ values, they are not included
in Fig.~\ref{F3}b because their low energy crystal electric field
(CEF) levels already contribute to $C_m(T)/T$ around 1\,K with the
consequent deviation from the $1/T^Q$ dependence in the sc-pm
phase. Expect these two cases, all the compounds analyzed in this
work own a well defined doublet GS. Notably, the recently studied
Sc-doped YbCo$_2$Zn$_{20}$ \cite{Gegenw} system also fits into the
NFL scaling presented in Fig.~\ref{F3}a for 19$\%$ of Sc content,
with the record low value of $T_0 = 1.2$\,K, whereas the low
temperature contribution of the first excited CEF doublet is
reflected in the large $\gamma_{HT}=1.1$\,J/molK$^2$ term. This
supports the vicinity of the parent compound to a QC-point as
discussed in Ref.\cite{Gegenw}.

\begin{figure}[tb]
\begin{center}
\includegraphics[width=19pc]{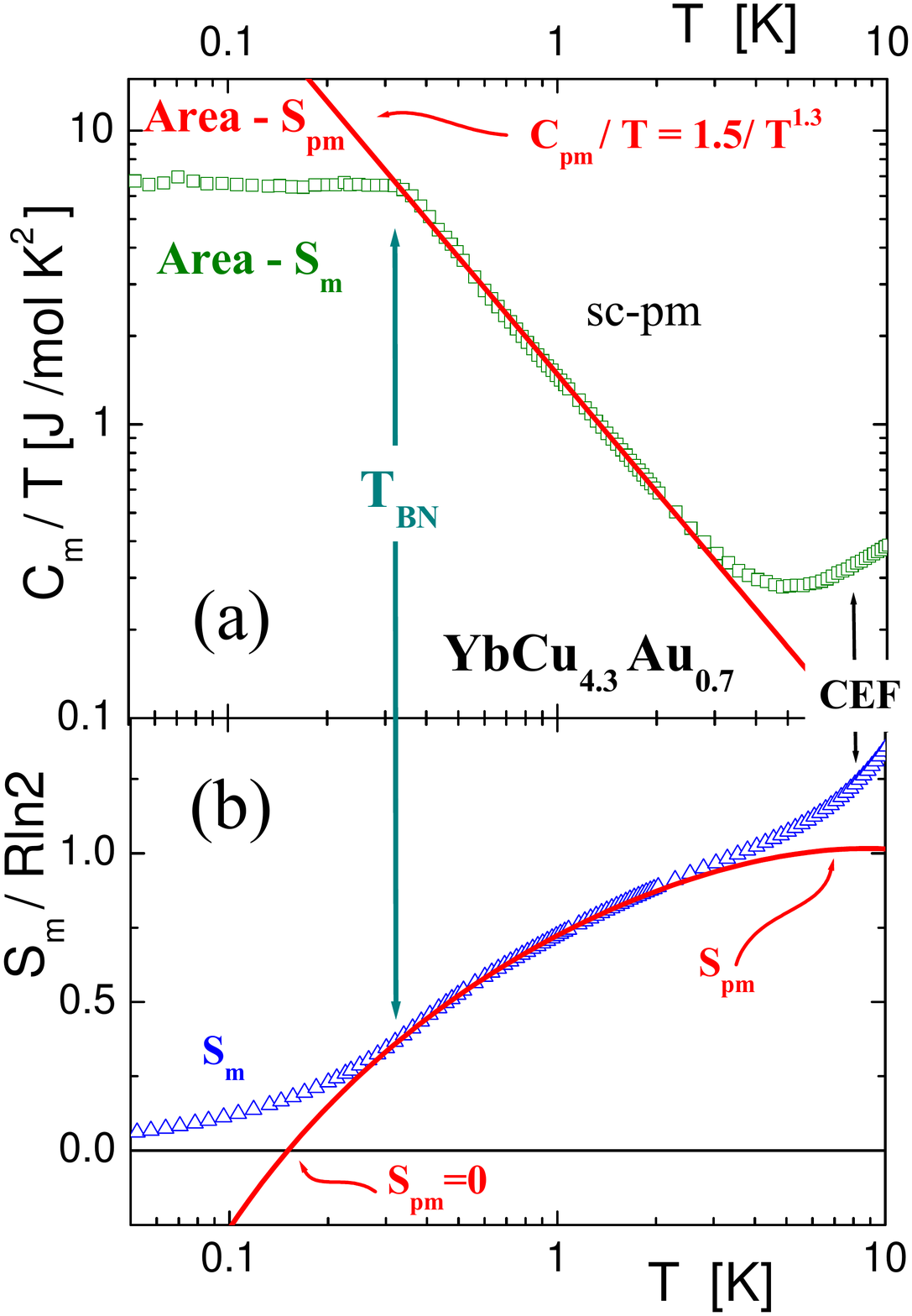}
\end{center}
\caption{(Color online) (a) Comparison between measured $C_m(T)/T$
and fitted $C_{pm}(T>T_{BN})/T$ dependencies in double logarithmic
representation for a magnetically frustrated system
\cite{YbCu5-xAux}. $S_m$ and $S_{pm}$ represent the areas from
which respective entropies are evaluated. (b) Thermal dependencies
of those entropies, showing how $S_{pm}\to 0$ at $T>0$. The value
of $S_{pm}(T\to \infty)= $R$\ln2$ is taken as reference for the
doublet GS. CEF indicates the excited crystal field levels
contribution above about 3\,K.} \label{F4}
\end{figure}

\section{Thermodynamic analysis of the $S_m(T)$ trajectory}

\subsection{Origin of the $C_m/T|_{T\to 0}$ upper limit: Entropy bottlenecks}

Since these systems are inhibited to order magnetically because of
their frustrated character, spin correlations develop remarkably
decreasing temperature in the sc-pm phase. As a consequence, the
density of magnetic excitations ($\propto C_{pm}/T$) grows
following a power law, as depicted in Fig.~\ref{F4}a for the
exemplary system YbCu$_{4.3}$Au$_{0.7}$, that extrapolates to a
mathematical singularity at $T = 0$. Therefore, a change of
behavior is expected at finite temperature (hereafter $T_{BN}$) in
order to escape such unphysical point. The question arises why
this change of regime occurs at certain characteristic temperature
$T_{BN}$ and how is its value established. To attempt to answer
these questions one should take into account the relevance of the
third law of thermodynamics in real systems at the $T\to 0$ limit.

In order to analyze how the third law (shortly expressed as
$S_m(T)|_{T\to 0} \geq 0$) intervenes in determining the ground
state (GS) of these VHF, the $C_m(T)/T$ dependence of the
magnetically frustrated YbCu$_{4.3}$Au$_{0.7}$ \cite{YbCu5-xAux}
is chosen. In its pyrochlore-type structure the Yb-magnetic
moments are located in tetrahedral vertices inhibiting magnetic
order due a 3D geometric frustration. This property guarantees
that no condensation of degrees of freedom occurs by magnetic
order induced via standard magnetic interactions. In fact, no
traces of order were detected in this system below $T_{BN}$ by
spectroscopic measurements \cite{Caretta}. In Fig.~\ref{F4}a one
may appreciate that $C_m(T>T_{BN} )/T$ increases obeying the power
law dependence $\propto 1.5/T^{1.3}$ which, below
$T_{BN}=350$\,mK, transforms into a 'plateau' with $C_m/T|_{T\to
0} \approx 6.5$\,J/mol\,K$^2$. This plateau was obtained after
subtracting the nuclear contribution of Yb atoms from the total
measured specific heat \cite{YbCu5-xAux}.

The area label as Area-$S_{pm}$ in Fig.~\ref{F4}a represents the
entropy computed as $S_{pm}=\int 1.5/T^{1.3} \times dT$ for an
heuristic system which does not change its $C_{pm}/T$ dependence
at $T_{BN}$ and clearly exceed the 'R$\ln2$' physical limit for a
doublet GS. On the contrary, the entropy extracted from the
measurement as $S_m=\int C_m/T \times dT$, represented by
Area-$S_m$, reveals that the diverging increase of
$C_{m}(T>T_{BN}/T$ towards low temperature runs across an 'entropy
bottleneck' (BN) \cite{Elsevier} that compels it to change
trajectory. This means that the change of $C_m(T)/T$ at $T=T_{BN}$
occurs because the system is constraint to not overcome the
$S_m=R\ln2$ value.

The comparison between $S_{pm}(T)$ and $S_{m}(T)$ trajectories is
included in Fig.~\ref{F4}b. Since this analysis corresponds to a
doublet GS, the $R\ln2$ upper limit at high temperature is taken
as the reference value. Notice that the contribution of the exited
CEF levels to $C_{m}(T)$ ($S_{m}(T)$) only occurs above about
$T=3$\,K. This comparison shows that both entropy trajectories
match very well within the $0.3 \geq T \geq 2$\,K range. However,
while $S_m(T<0.3\,K)$ turns pointing to the expected $S_m|_{T\to
0} = 0$ limit, $S_{pm}(T)$ becomes zero a finite temperature (at
$T\approx 0.17$\,K in this case) and keeps decreasing into
negative values. It is clear that below $T=T_{BN}$ the third law
imposes another trajectory to $S_m(T)$, which departs from
$S_{pm}(T)$ in order to not overcome the physical limit of
$S_m|_{T\to 0}= 0$. A relevant aspect of this change of trajectory
for $S_m(T)$ arises from the fact that it is not driven by
magnetic interactions but by the thermodynamic constraint:
$S_m|_{T\to 0}\geq 0$.

\begin{figure}[tb]
\begin{center}
\includegraphics[width=20pc]{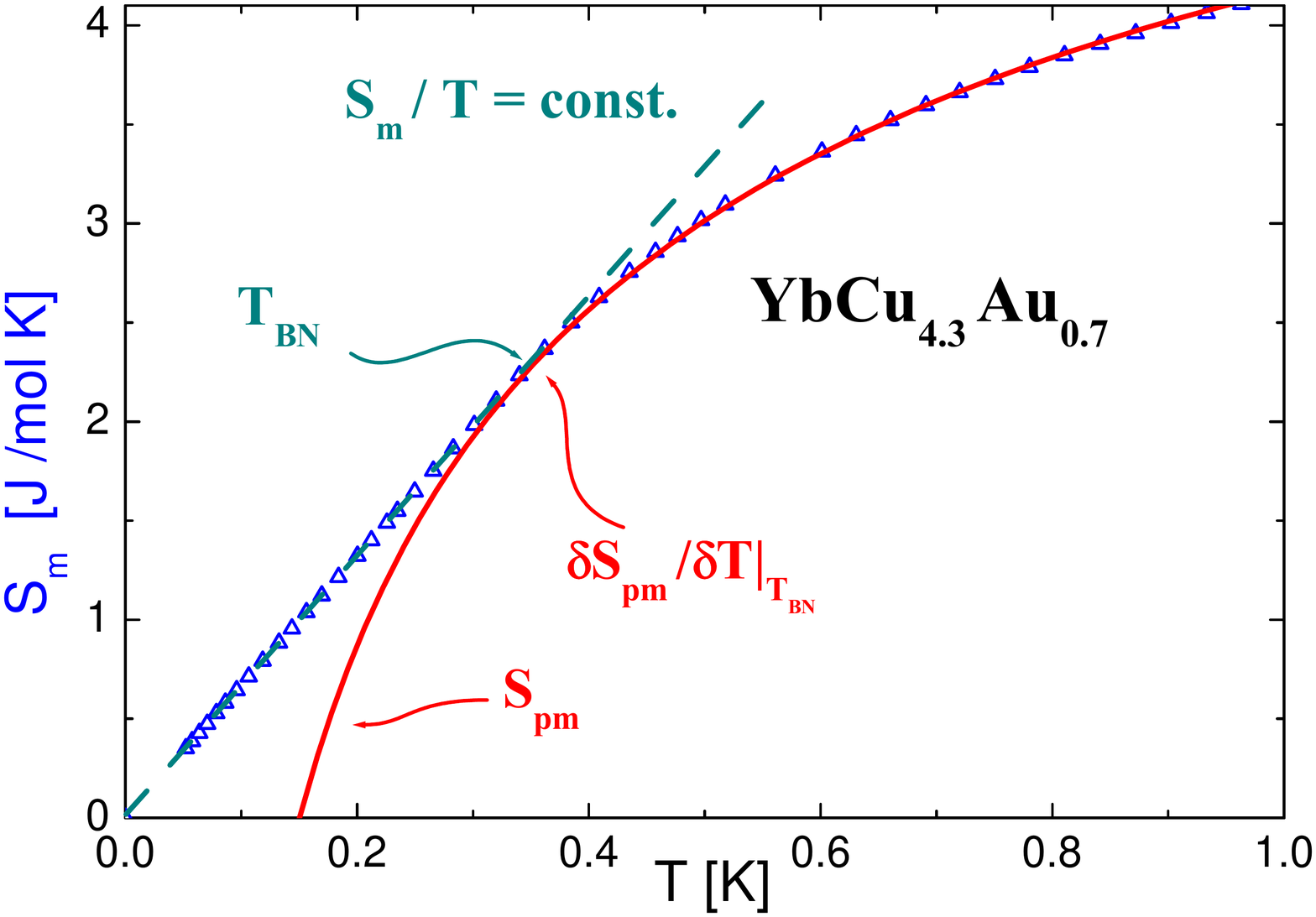}
\end{center}
\caption{(Color online) Analysis of the thermodynamic condition
producing the entropy deviation from the sc-pm trajectory.
Continuous curve: $S_{pm}(T)$, dashed line: $S_m/T =$ constant.}
\label{F5}
\end{figure}

\subsection{Conditions to deviate from the sc-pm behavior}

Once the origin of the change of regime was discussed, one should
trace the thermodynamic conditions able to split $S_m(T)$ and
$S_{pm}(T)$ trajectories at $T=T_{BN}$. For such a purpose we
focus on $S_m(T_{BN})$ and $S_{pm}(T_{BN})$ coincidences. Looking
at the same frustrated system YbCu$_{4.3}$Au$_{0.7}$, one may see
in Fig.~\ref{F5} that at $T_{BN}$ the entropy derivative $\partial
S_{m} /\partial T$ ( = $\partial S_{pm} /\partial T$) coincides
with the $S_m(T<T_{BN})/T$ ratio that extrapolates to $S_m|_{T\to
0} = 0$. Then, taking into account that $\partial S_m /\partial T
\equiv C_m/T$, one finds that $C_m/T_{BN}= S_m/T_{BN}$ and
therefore $S_m = C_m$ at that temperature.

Since the specific heat is defined as $C_m \equiv \partial E_m
/\partial T$, where $E_m$ is the magnetic enthalpy, the $S_m =
C_m$ equality can be written as: $S_m - \partial E_m /\partial T =
0$. This expression coincides with the Planck's potential: $\Phi =
S - E/T$ \cite{Planck} provided that $\partial E_m /\partial T =
E_m/T$, which is the case of the so-called 'plateau' regime of the
VHF showing $C_m/T|_{T\to 0} =$ const. That property fulfils the
$\Phi = S_m - E_m /T = 0$ condition. Another relationship that
characterizes this change if regime can be extracted from
$\partial S_m /\partial T = S_m/T$ writing $\partial S_m/S_m =
\partial T/T$. This implies that at $T = T_{BN}$ entropy and
temperature (i.e. the thermal energy) change with the same ratio
like in a sort of Grueneisen ratio: $\partial \ln{S_m} / \partial
\ln T = 1$.

\subsection{Other cases with similar thermal $C_{pm}/T$ dependencies}

\begin{figure}[tb]
\begin{center}
\includegraphics[width=20pc]{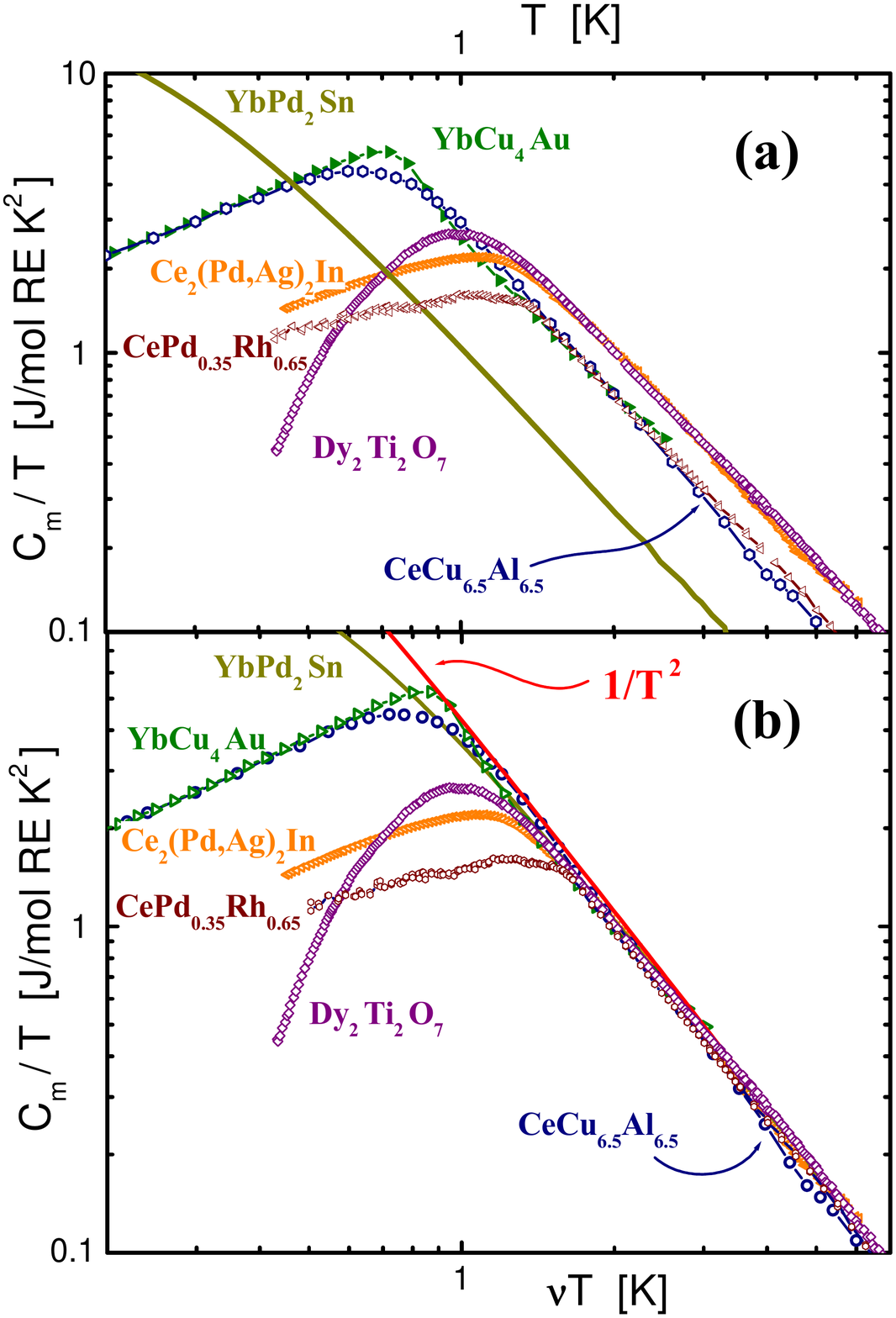}
\end{center}
\caption{(Color online) (a) Compounds with $C_m(T)/T$ maxima at
$T\approx 1$\,K and the same power law dependence in their 'sc-pm'
phases, after \cite{RefFig6,Micha}. Notice the double logarithmic
representation and the mass normalization to a single rare earth
(RE) atom. (b) Same compounds scaled by normalizing their
respective temperatures with a factor $\nu$, taking
Dy$_2$Ti$_2$O$_7$ and Ce$_2$(Pd$_{0.5}$Ag$_{0.5}$)$_2$In as
reference. A $1/T^2$ function is included to show the coincident
dependence of these compounds.} \label{F6}
\end{figure}

In order to recognize whether this 'entropy bottleneck' effect
only occurs in the VHF with a $C_m/T|_{T\to 0}$ 'plateau' below
$T_{BN}$ or it is a general property, the same analysis is applied
to a number of compounds collected in Fig.~\ref{F6}a which are
inhibited to order magnetically. The figure contains some
geometrically frustrated cases, like the 3D-pyrochlores YbCu$_4$Ni
and Dy$_2$Ti$_2$O$_7$ spin-ice \cite{Dy2Ti2O7} and the 2D
Ce$_{2.15}$(Pd$_{0.5}$Ag$_{0.5}$)$_{1.95}$In$_{0.9}$
\cite{CePdAgIn}, hereafter quoted as
Ce$_2$(Pd$_{0.5}$Ag$_{0.5}$)$_2$In for simplicity. The latter
system shows on-plane triangular coordination between
Ce-nearest-neighbors in the Mo$_2$B$_2$Fe-type structure. A
striking feature observed in all the mentioned systems is the
nearly coincident power law dependence $C_m/T \propto 1/T^Q$ in
the respective 'sc-pm' phases.

To explore the ampleness of this peculiar dependence besides
frustration phenomena, other systems like: YbPt$_2$Sn
\cite{YbPt2Sn}, CeCu$_{6.5}$Al$_{6.5}$ \cite{CeCu65Al65} and
CePd$_{0.35}$Rh$_{0.65}$ single crystal \cite{Micha} are included
into this comparison because they show the same $C_m/T \propto
1/T^2$ dependence. Although the competition between the
first-$J_1$ and second-$J_2$ neighbor magnetic exchange
interactions may cause frustration in simple structures, the case
of CePd$_{0.35}$Rh$_{0.65}$ seems to escape to that pattern
because it was actually investigated in the context of a QC regime
\cite{CePdRh}. This supports one of the main points of the present
analysis: the change in the $S_m(T)$ trajectory is due to
thermodynamic constraints independently of the reason why magnetic
order is inhibited. In other words, the lack of magnetic order
down to very low temperature provides the conditions for the
entropy-bottleneck occurrence.

Besides the coincident property of all systems included in
Fig.~\ref{F6}a, that their $C_m(T>T_{BN})/T$ dependencies is
described with the same heuristic formula $C_{pm}/T = D/(T^Q+E)$
\cite{SerHvL} with $Q=2\pm 0.1$, one finds that Dy$_2$Ti$_2$O$_7$
and Ce$_2$(Pd$_{0.5}$Ag$_{0.5}$)$_2$In also coincide in all their
fitting parameters: $D=4.5$\,J/mol at. and $E=0.3$\,K$^2$
\cite{CePdAgIn}. These systems do not show the VHF-plateau
behavior because once reached the entropy-bottleneck conditions
they access to some alternative minima of their free energy, which
is evidenced by the decrease of their $C_m/T|_{T<T_{BN}}$.

As a consequence of their common power law dependence, these
compounds can be scaled by a simple normalization of their
$D_{\nu}$ coefficients: $D_{\nu} = \nu D$, taking as (arbitrary)
reference the coincident $D=4.5$\,J/mol\,RE. value of
Dy$_2$Ti$_2$O$_7$ and Ce$_2$(Pd$_{0.5}$Ag$_{0.5}$)$_2$In, see
Fig.~\ref{F6}b. In this comparison, the $C_m(T)/T$ properties of
YbPt$_2$Sn merit some comment because it coincides with some
characteristics of this group but it differs from others. Together
with its homologous YbPt$_2$In \cite{YbPt2Sn}, its $C_{pm}(T)/T
\propto 1/T^2$ dependence coincides with the common $Q\approx 2$
exponent of other compounds despite showing the lowest temperature
maximum of $C_m(T)/T$. This feature supports the fact that the
underlying mechanism responsible for the entropy-bottlenecks
formation does not depend of an energy scale but reflect a general
thermodynamic property. Besides that, these YbPt$_2$X compounds
show that the $\approx 7$\,J/molK$^2$ value observed in the
'plateau' group is not an upper limit for $C_{m}/T|_{T\to 0}$ but
a characteristic of VHF.

\begin{figure}[tb]
\begin{center}
\includegraphics[width=20pc]{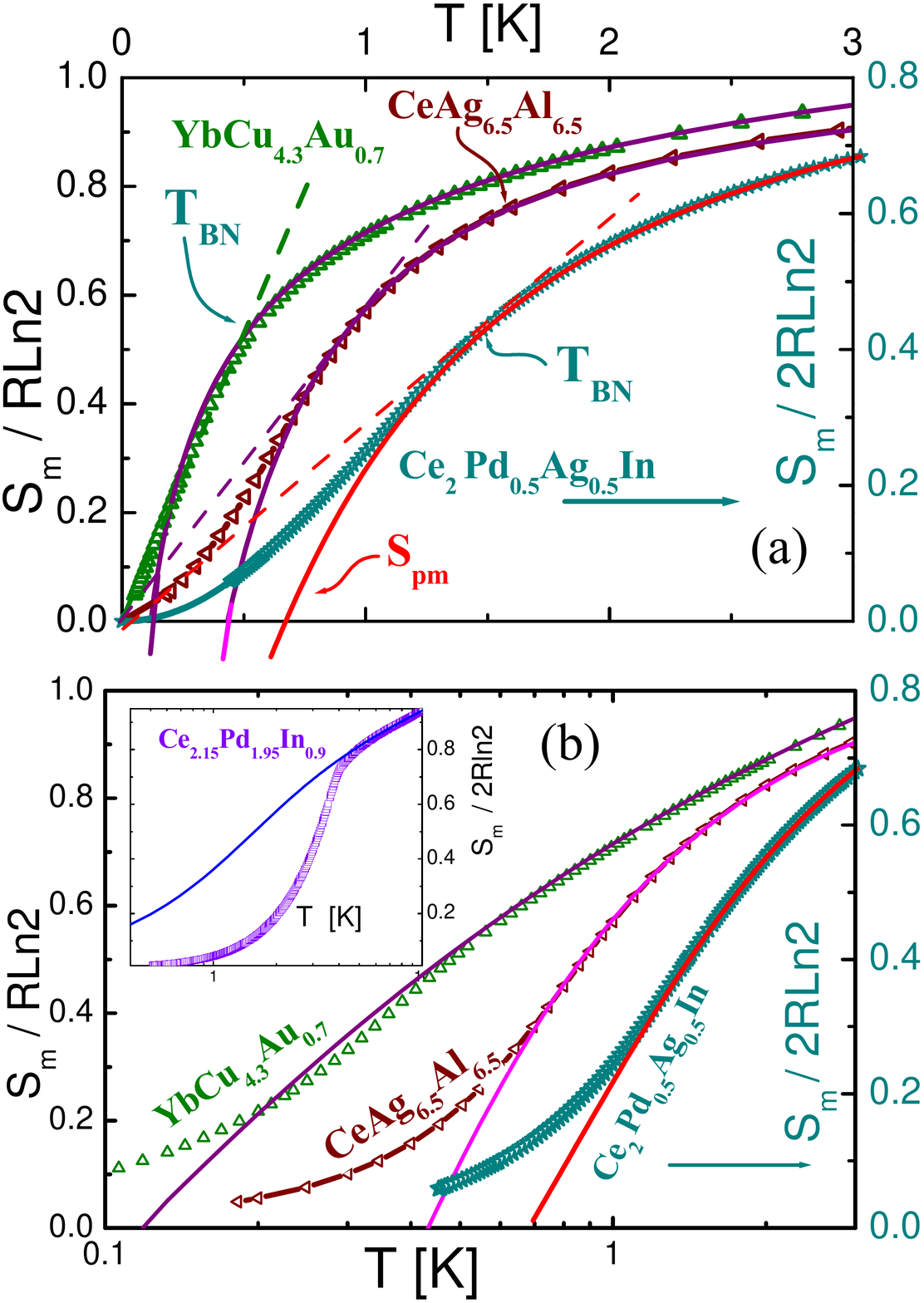}
\end{center}
\caption{(Color online) a) $S_m(T)$ dependencies of three
compounds showing their respective $S_{pm} = 0$ extrapolations at
$T>0$ and respective $S_m/T$ (dashed) lines defining $T_{BN}$. b)
Same entropies dependencies in a log(T/K) representation compared
(see the inset) with that of a standard ferromagnetic compound.}
\label{F7}
\end{figure}

The coincident $Q\approx 2$ value observed in the power law
thermal dependence of the five rare earth based compounds
collected in Fig.~\ref{F6} clearly suggests that some common
physics underlies in their 'sc-pm' behavior. Although power law
dependencies for $C_m/T$ are frequently reported in model
predictions, the exponents use to be much smaller \cite{Stewart01}
than the observed. Nevertheless, a $Q=2$ exponent was reported
\cite{Schofield} to describe the thermal properties of Kondo type
systems assuming a uniform distribution of $T_K$ between 0 and a
cut-off temperature with a $C_m/T \propto 1/(T^2+T_K^2)$
dependence. However, mimicking $T_K$ as the spin fluctuation
energy of the frustrated-moment orientation in a sort of
spin-liquid scenario, that expression nicely fits the reported
experimental results.

\begin{figure}[tb]
\begin{center}
\includegraphics[width=20pc]{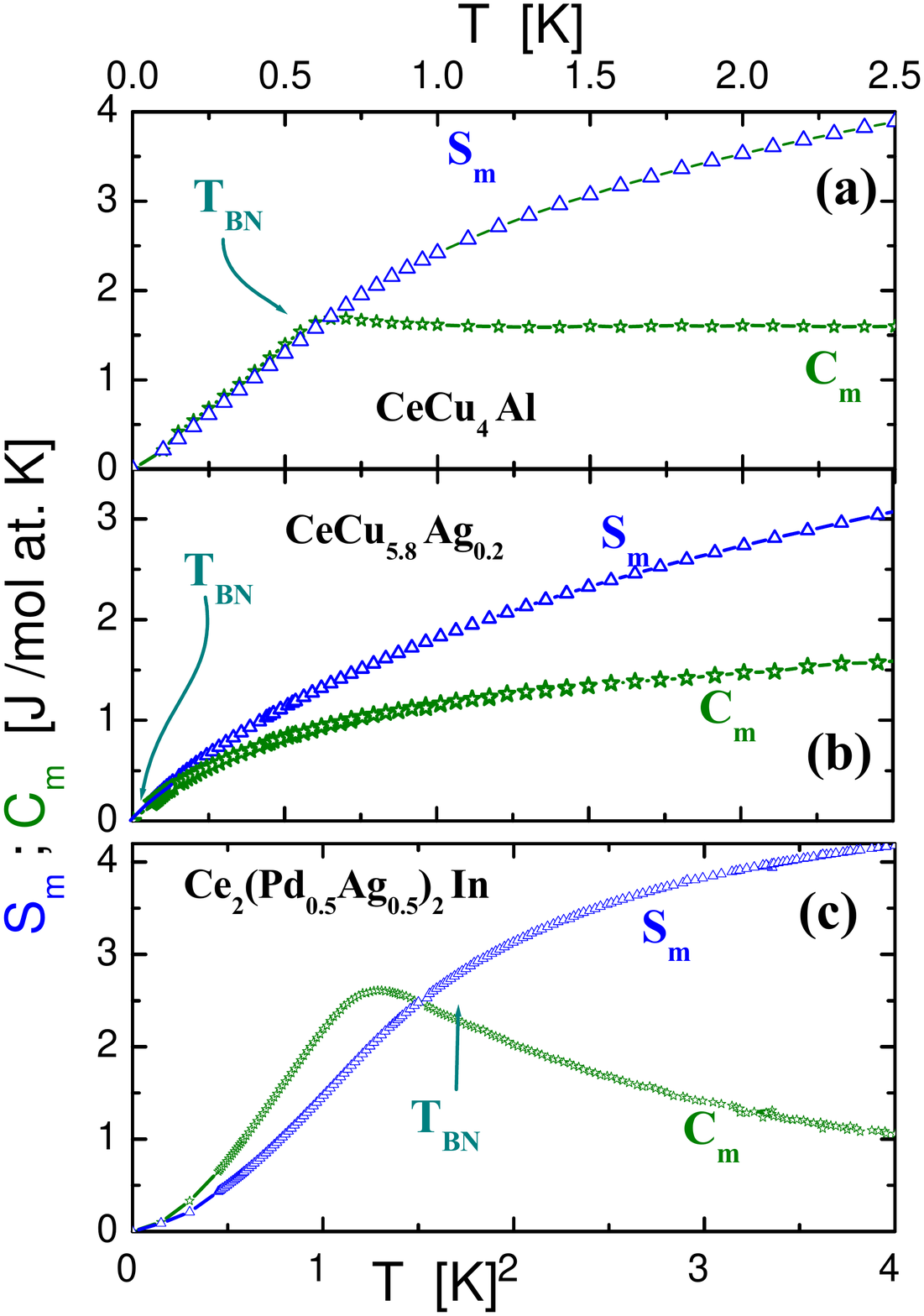}
\end{center}
\caption{(Color online) Comparison of $S_m(T)$ and $C_m(T)$
dependencies for three different types of GS: (a) a 'plateau' type
with $T_{BN} \approx 0.6$\,K, data after \cite{CeCu4Al},(b) the
case of a NFL \cite{CeCu58Ag02}, where $\Phi(T) = 0$ occurs at
$T_{BN}=0$, and (c) a 2D frustrated system showing a
$C_m(T)/T_{max}$ at $\approx 1$\,K, data after \cite{CePdAgIn}.}
\label{F8}
\end{figure}

\subsection{Comparison between different entropy behaviors}

Regarding the thermal dependence of the entropy, the same
discussion performed in Fig.~\ref{F4} about the change of
trajectory at $T_{BN}$ can be applied to these compounds, see
Fig.~\ref{F7}a. Two representative cases are included in the
figure: Ce$_2$(Pd$_{0.5}$Ag$_{0.5}$)$_2$In and
CeCu$_{6.5}$Al$_{6.5}$, and compared with YbCu$_{4.3}$Au$_{0.7}$
already analyzed in Fig.~\ref{F4} as a member of the 'plateau'
group. Each $S_{pm}(T)$ dependence is computed as $S_{pm} = \int
C_{pm}/T \times dT$, with respective $C_{pm}(T>T_{BN})/T = 1.46
/T^{1.3}$ for YbCu$_{4.3}$Au$_{0.7}$, $= 9.1 / (T^2 + 0.35)$ for
Ce$_2$(Pd$_{0.5}$Ag$_{0.5}$)$_2$In and $= 3/T^{2}$ for
CeCu$_{6.5}$Al$_{6.5}$. As it can be appreciated, all these
$S_{pm}(T)$ trajectories cross the $S_m =0$ axis at finite
temperature because $S_{pm}|_{T\to 0}$ extrapolates to the
unphysical value $S_{pm}|_{T= 0}< 0$.

Focusing to the actual value of $S_m$ at $T_{BN}$, it is evident
from Fig.~\ref{F7}a that $S_m(T_{BN}) \approx 1/2 R\ln2$. Within a
significantly low dispersion, this observation includes the
CePt$_2$X compounds \cite{YbPt2Sn} and frustrated systems like
Ce$_2$(Pd$_{0.5}$Ag$_{0.5}$)$_2$In and Dy$_2$Ti$_2$O$_7$.
Interestingly, the Dy$_2$Ti$_2$O$_7$ spin-ice fits better into
this systematic once the Pauling's residual entropy
$S_0=(1/2)R\ln(3/2)$ \cite{RamirezNat} is included into the
'total' $S_{tot} = S_m + S_0$.

In Fig.~\ref{F7}b the same picture is shown in a 'log(T/K)'
representation in order to better distinguish the detachment
between measured $S_m(T)$ and computed $S_{pm}(T)$ below
$T=T_{BN}$. Interestingly, in YbCu$_{4.3}$Au$_{0.7}$ the $S_m/T$
slope is slightly lower than the $\partial S_{pm}/\partial T$
around $T_{BN}$ like as a reminiscence of a second order
transition. Such a small effect is within the scale of eventual
composition inhomogeneities at the surface of these
poly-crystalline samples. As an inset in Fig.~\ref{F7}b, the case
of the ferromagnet Ce$_{2.15}$Pd$_{1.95}$In$_{0.9}$
\cite{Ce2Pd2In} is included for comparison. This magnetically
ordered compound was selected because it shows a well defined
$C_m(T_C)$ jump at relatively low temperature $T_C = 4.1$\,K. It
is evident that in this case the corresponding paramagnetic
$S_{pm}(T>T_C)$ extrapolation below $T_C$ remains above the
measured $S_m(T<T_C)$ values, in clear contrast with the not
ordered compounds. One should notice that the $\partial^2
S_m/\partial T^2$ derivative show an inflection point at
$T=T_{BN}$ whereas at $T=T_C$ there is a discontinuity.

To check the applicability of the $C_m=S_m$ equality in a wider
range of behaviors it is illustrative to compare the $C_m(T)$ and
$S_m(T)$ dependencies in other systems that do not order
magnetically but having GS of different nature. In Fig.~\ref{F8}a
the stoichiometric compound CeCu$_{4}$Al with a $T_{BN} \approx
0.6$\,K, \cite{CeCu4Al} is included as another example showing
$C_m/T|_{T\to 0}$ constant and thence a coincident $C_m(T)$ and
$S_m(T)$ below $T_{BN}$. Interestingly, as it is depicted in
Fig.~\ref{F8}b the NFL CeCu$_{5.8}$Ag$_{0.2}$ \cite{CeCu58Ag02}
reaches the $C_m(T) = S_m(T)$ equality at $T=0$ with consequent
$T_{BN}=0$. Due to their $C_m \propto - T \times \ln (T/T_0)$
dependencies the $C_m(T)- S_m(T)$ difference is expected to
increase linearly with temperature. CeCu$_{5.8}$Ag$_{0.2}$ was
chosen as a NFL exemplary system because it shows the largest
measured value $C_m/T=3$\,J/mol\,K$^2$ at $T=60$\,mK, see
Fig.~\ref{F2}. In spite of this, it does not reach the values of
VHF even at $T\to 0$ because it shows a slight downwards deviation
that extrapolates to $C_m/T|_{T=0} < 4$\,J/mol\,K$^2$.
Fig.~\ref{F8}c includes an example of a 2D frustrated system
Ce$_2$(Pd$_{0.5}$Ag$_{0.5})$In \cite{CePdAgIn}. As expected from
the analysis performed in Fig.~\ref{F7}a, the $C_m=S_m$ equality
occurs where $C_m(T)/T$ deviates from the sc-pm regime described
by the $\propto 1/T^2$ dependence. This temperature is close but
not exactly at the maximum of $C_m(T)$. A replica of this behavior
is obtained for the Dy$_2$Ti$_2$0$_7$ spin-ice. This comparison
supports the conclusion that the entropy-bottleneck is a general
effect occurring in systems not able to reach magnetic order.

\section{Thermodynamic behavior at $T\leq T_{BN}$}

In order to gain insight into the peculiar behavior of these
systems, one should focus on the two main questions arising from
the observed phenomenology that mostly concerns the change of
regime at $T=T_{BN}$ and the nature of the ground state beyond the
entropy-bottleneck. Although the usual Kondo scenario, able to
explain the lack of magnetic order can be discarded in these
systems \cite{KondoCePdRh}, in the range of temperatures where the
entropy-bottlenecks occur ($T_{BN}\leq 1$\,K) quantum fluctuations
play an important role in the thermodynamic equilibrium because
they may overcome thermal fluctuations \cite{SerHvL}.

According to a recent $'QK'$ phase diagram which includes
frustration as a tuning parameter ($'Q'$), competing with Kondo
effect ($'K'$) to define the GS of HF materials \cite{Coleman},
the frustrated systems analyzed in this work can be placed on the
'spin liquid' (SL) side. Although that schematic phase diagram was
proposed for antiferromagnetic (AFM) order (specifically related
to Shastry-Shutherland lattice formation) as a wider concept it
might include any possible type of order parameter. The fact that
Ce$_2$(Pd$_{0.5}$Ag$_{0.5}$)$_2$In \cite{CePdAgIn} is related to
the Shastry-Shutherland lattice behavior of Ce$_2$Pd$_2$Sn
\cite{Ce2Pd2Sn} indicates that different GS of similar energies
may become accessible, being the AFM order one of those
possibilities. One should remind that AFM-$J_R$ interactions and
geometrical factors are basic ingredients to inhibit magnetic
order through frustration effects.

\subsection{Characteristics of the $T_{BN}$ transition}

\begin{figure}[tb]
\begin{center}
\includegraphics[width=21pc]{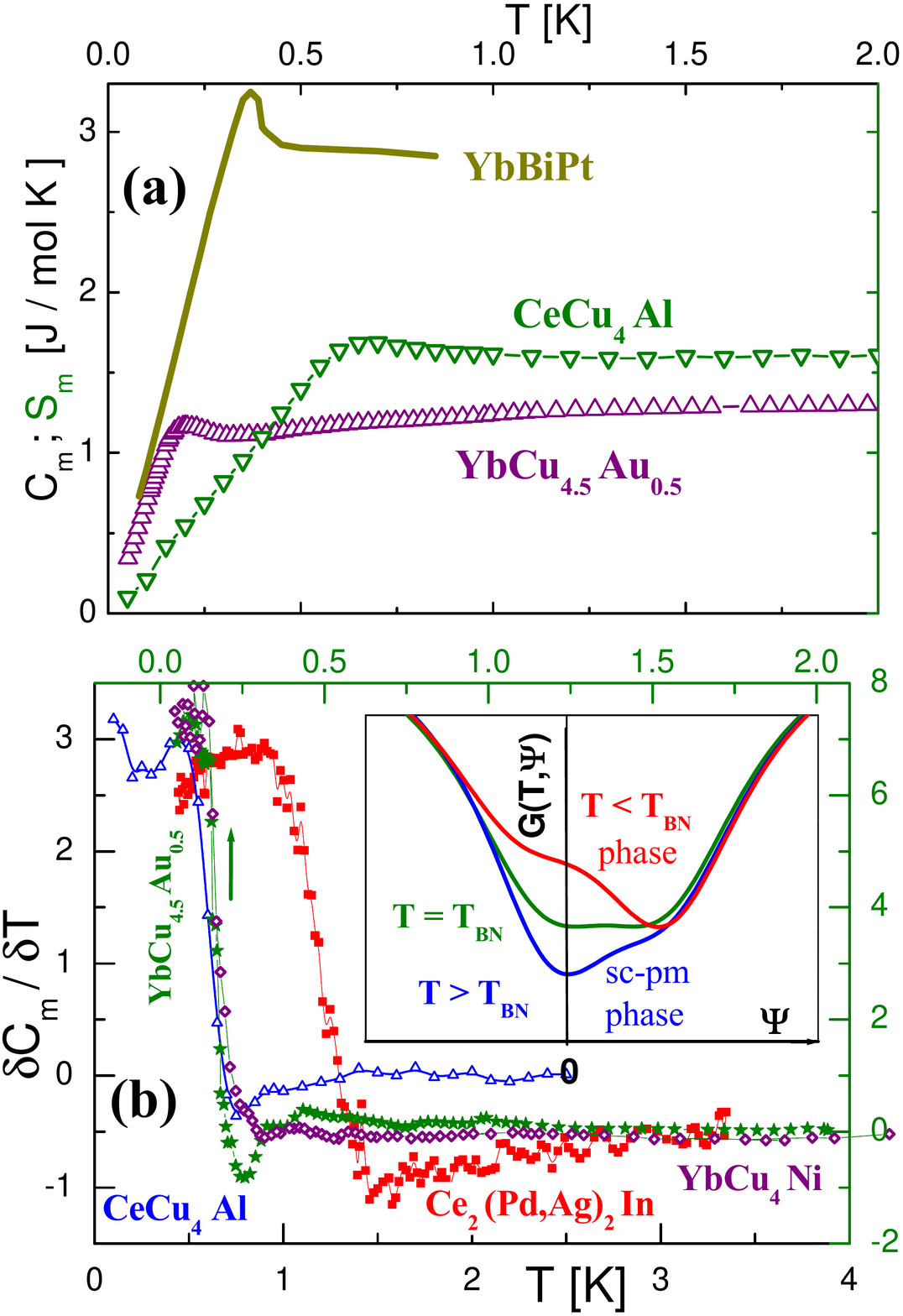}
\end{center}
\caption{(Color online) (a) Three illustrative cases: YbBiPt
\cite{YbBiPt}, CeCu$_4$Al \cite{CeCu4Al} and Yb
Cu$_{4.5}$Au$_{0.5}$ \cite {YbCu5-xAux}, showing how $C_m(T\to 0)>
0$ before the entropy-bottleneck in reached. (b) $\partial
C_m/\partial T$ derivative of some studied compounds showing a
discontinuity at $T=T_{BN}$. Inset: isothermal $G(\Psi)$
representations for three different temperatures including a broad
minimum at $T=T_{BN}$, after \cite{Atoms58}.} \label{F9}
\end{figure}

In Fig.~\ref{F9}a the $C_m(T)$ dependence of other three systems
are presented to show the effect of the entropy-bottleneck through
an alternative picture. From the $C_m(T>T_{BN}) \approx$ const.
behavior, it is evident that it would extrapolate to a
non-physical $C_m \neq 0$ at $T=0$ unless a change trajectory
occurs once the $S_m=C_m$ equality is reached. The small jump at
$T_{BN}$ was already attributed to possible grains-surface
contribution in poly-crystalline samples.

Since the change of regime occurs without a $C_m(T)$ jump, the
following scenario can be proposed: the minimum of the free energy
of the sc-pm phase ($G_{pm}$) blurs out at $T_{BN}$ because it is
not workable anymore and the system is compelled to slide into any
other energetically accessible $G(T)$-minimum. This description is
schematically depicted in the inset of Fig.~\ref{F9}b
\cite{Atoms58} as a $G_{pm}(\Psi)$ dependence on the order
parameter $\Psi$ at different temperatures. This continuous creep
along the $G(\Psi,T)$ surface is related to a higher order
discontinuity of the $G(T)$ derivative, i.e. $\partial
G^3/\partial^3T$, that emerges as a discontinuity in $\partial
C_m/\partial T$ of third order transition character
\cite{Pippard}, clearly fulfilled by the transitions collected in
Fig.~\ref{F9}b.

Another evidence for the third order character of this transition
is given by the entropy trajectory in the low temperature phase.
As it is shown in the inset of Fig.~\ref{F7}b for the case of a
model second order FM transition, $S_m(T)<S_{pm}(T)$ below $T_C$
because a magnetic order parameter $\Psi(T<T_C)$ develops
condensing the entropy extrapolated from $T>T_C$. On the contrary,
in the systems presented in Fig.~\ref{F7}, $S_m(T)>S_{pm}(T)$
below $T_{BN}$. This difference is a direct consequence of the
$\partial S_m^2/\partial^2 T$ discontinuity at $T=T_C$ respect to
the inflection point at $T=T_{BN}$.

\subsection{Nature of the ground state beyond the entropy bottleneck}

\begin{figure}[tb]
\begin{center}
\includegraphics[width=19pc]{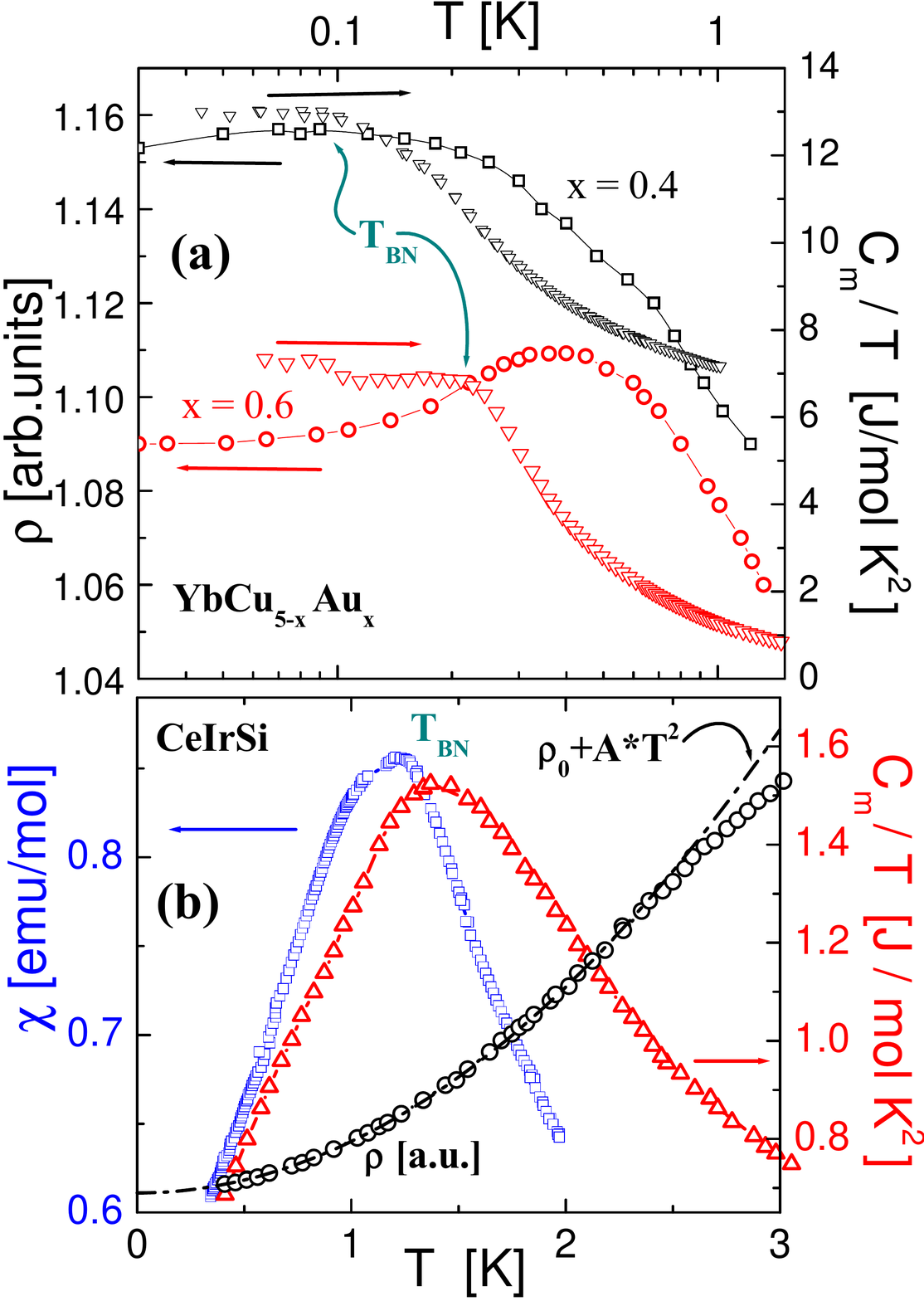}
\end{center}
\caption{(Color online) Continuous formation of a coherent GS
indicted by electrical resistivity of systems belonging to: (a)
the group showing a $C_m/T|_{T\to 0}$ 'plateau'
\cite{YbCu5-xAux,ResYbCu5Au} and (b) to those with the $C_m/T$
maximum at $T_{BN}\approx 1$\,K, after \cite{Kneidinge}.}
\label{F10}
\end{figure}

Despite of the scarce thermodynamic information available at very
low temperature, some relevant features can be observed in the
systems showing the $C_m/T|_{T\to 0} \approx 7$\,J/molK$^2$
'plateau'. Among them, the already mentioned absence of magnetic
order confirmed in YbCu$_{5-x}$Au$_x$ down to 20\,mK using $\mu$SR
and NQR techniques \cite{Caretta} is significant. Coincidentally,
coherent electronic scattering is evident in $\rho(T)$ from
$T>T_{BN}$ in samples with $x=0.4$ and 0.6 \cite{ResYbCu5Au}. This
behavior is compared with the establishment of the specific heat
'plateau' at $T_{BN}$, see Fig.~\ref{F10}a.

Similar $\rho(T)$ coherence effects are observed in PrInAg$_2$
around $T_{BN}\approx 250$\,mK \cite{PrInAg2}. Interestingly, the
doublet GS of this compounds is not of Kramer's character because
of the integer $J=4$ total angular moment of Pr. This indicates
that the type of GS wave function is irrelevant in the formation
of the entropy-bottlenecks, supporting the purely thermodynamic
origin of this phenomenon. Also the magnetic susceptibility
($\chi$) of YbCo$_2$Zn$_{20}$ confirms the lack of magnetic order
below $T_{BN}$ because its $\chi(T)$ dependence simply shows a
broad maximum at $T\approx 300$\,mK \cite{Takeuchi} and the
formation of a coherent lattice state well above $T_{BN}$
\cite{Gegenw}.

From these experimental evidences one may conclude that the GS of
these compounds behave more likely as Fermi or Spin Liquids rather
than long range $J_R$-like interacting moments. This means that
the change occurring at $T_{BN}$ is not driven by standard
magnetic interactions. Taking into account that this phenomenon
occurs at the mK range of temperature, dominated by quantum
fluctuations, eventual tunnelling effect between states of
equivalent energy may act as hopping mechanism. In that case the
formation of a very narrow band-like of excitations, reflected in
a very high and constant $C_m/T|_{T\to 0}$ value, may arise.
However, to our knowledge, there is no specific experimental
indication in the literature for any of these scenarios.

The absence of magnetic order was also confirmed in the exemplary
spin-ice for the family of compounds showing a $C_m(T)/T$ anomaly
at $T_{BN}\approx 1$\,K by transverse-field $\mu$SR experiments in
Dy$_2$Ti$_2$O$_7$ \cite{Dunsiger}. A coherent GS formation is also
supported by $\rho(T)$ measurements in CeIrSi \cite{Kneidinge},
where Ce-atoms coordination provides the geometrical condition for
magnetic frustration. In this compound $\rho(T)$ obeys a $\rho =
\rho_0+A \times T^2$ dependence for $T\leq 2.4$\,K which reveals a
coherent electronic scattering, see Fig.~\ref{F10}b. This range of
temperature fully covers respective $C_m(T)/T$ and $\chi(T)$
anomalies around $T_{BN} \approx 1.4$\,K. The absence of magnetic
order was also remarked in YbBiPt \cite{YbBiPt}, with a
considerably small $T_K$ and a monotonous decrease of $\rho(T)$.

\subsection{Analysis of the thermal dependencies of the entropy at
$T\to 0$ in real systems}

\begin{figure}[tb]
\begin{center}
\includegraphics[width=19pc]{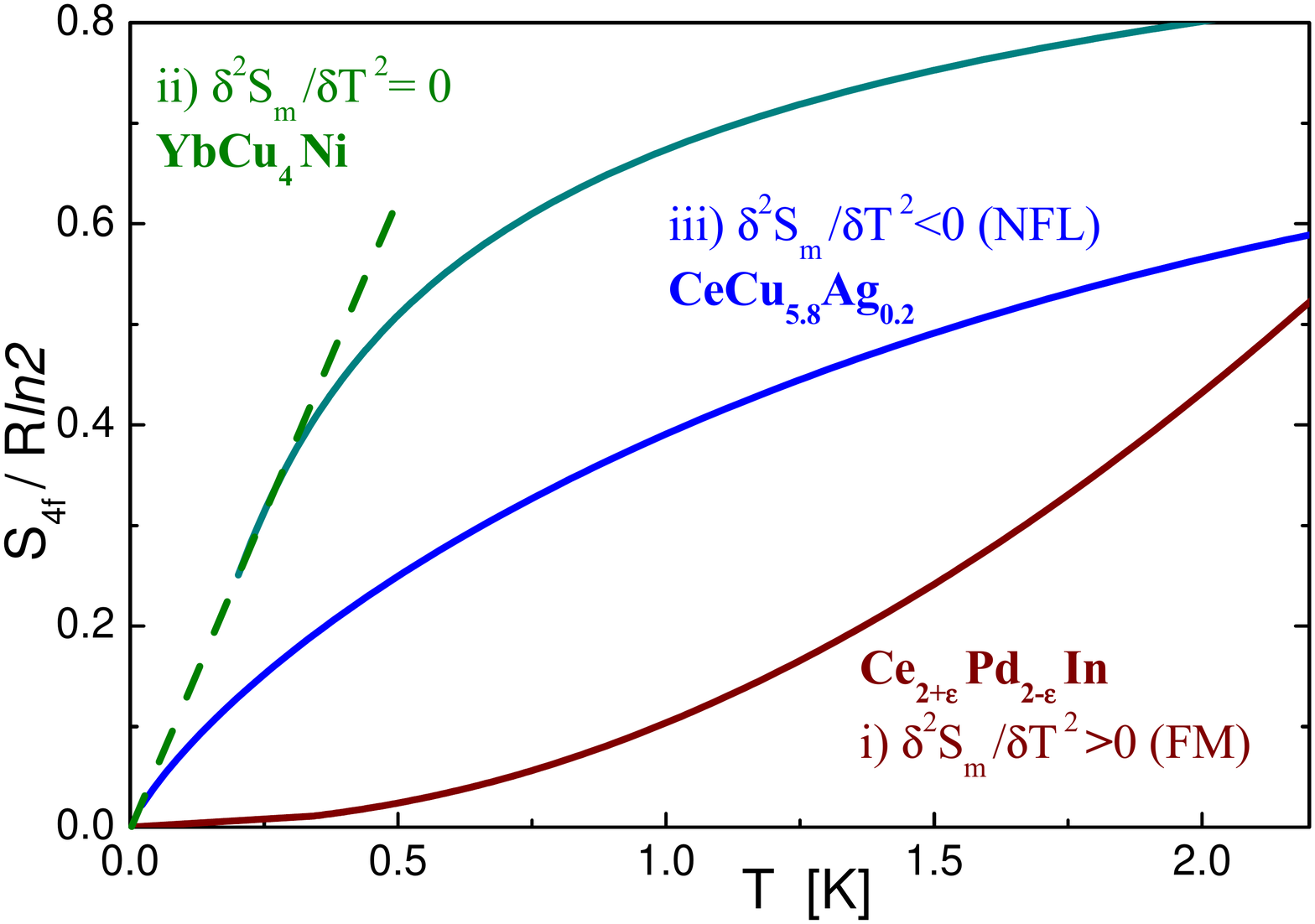}
\end{center}
\caption{(Color online) Different ways for $S_m|_{T\to 0}$
according to its second derivative $\partial^2 S_m/\partial T^2$.
}\label{F11}
\end{figure}

Despite of the leafy literature devoted to the $S_m|_{T\to 0}\geq
0$ postulate, there is scarce information about the $S_m(T)$
derivatives approaching $T=0$ \cite{Abriata}. Although the
unattainability of that limit prevents any experimental
approximation, the present knowledge of the $S_m(T)$ behavior of
Ce- and Yb-based compounds not showing magnetic order allows to
explore that field from a phenomenological point of view. This
topic is becoming relevant because the efficiency of new
cryo-materials for adiabatic demagnetization at very low
temperature depends on how magnetic field affects the $S_m(T)$ and
its slope, i.e. $C_m/T = \partial S_m\partial T$, at the proximity
of $T=0$.

Some interesting features can be extracted from the analysis of
the possible trajectories of $S_m|_{T\to 0}$, schematically drown
in Fig.~\ref{F11} and formerly discussed in Ref.\cite{PhilMag}.
The three possible curvatures presented in the figure can be
classified according to their second thermal derivatives
$\partial^2 S_m/\partial T^2$: i) the positive ($>0$) curvature
corresponds to standard ordered systems acceding to a cooperative
GS, represented by the FM compound
Ce$_{2+\epsilon}$Pd$_{2-\epsilon}$In \cite{Ce2Pd2In}, ii) the zero
curvature ($ = 0$) identifies to the group showing a 'plateau' in
$C_m/T|_{T\to 0}$ and corresponds to a continuous density of
excitations, like e.g. in FL. iii) The negative curvature ($<0$)
is characteristic of frustrated systems in the 'sc-pm' regime,
which cannot extrapolate to $T=0$ because it would end at a
singularity. As a consequence, no experimental example can be
quoted since real systems are compelled to modify their $S_m(T)$
trajectory at $T_{BN}$. The lowest temperature $\partial^2
S_m/\partial T^2|_{T\to 0}<0$ curves are observed in NFL compounds
because their $C_m\propto - T\times \ln(T/T_0)$ dependence
extrapolate to $C_m|_{T\to 0}=0$, see the case of
CeCu$_{5.8}$Ag$_{0.2}$ \cite{CeCu58Ag02} in Fig.~\ref{F8}b. As
mentioned before, this case corresponds to a $T_{BN} = 0$
scenario. The moderate increase of the entropy at the mK range
produced by the extended energy range of magnetic excitations,
$S_m(T) \propto T - T\times ln(T/T_0)$, allows the closest
approach to $\partial^2 S_m/\partial T^2|_{T=0} < 0$. In
Fig.~\ref{F11}, CeCu$_{5.8}$Ag$_{0.2}$ is taken as the exemplary
system with a pure doublet GS because, in the case of
(Yb$_{0.8}$Sc$_{0.2})$Co$_2$Zn$_{20}$ \cite{Gegenw}, the first
excited CEF level already contributes to $S_m(T)$ within the
temperature range of the figure. Interestingly, such Sc-doping is
able to drive the $T_{BN} \approx 200$\,mK of Yb$_2$Zn$_{20}$ down
to zero as atomic disorder weakens the spin-correlations that
reduces $C_m/T$ above $T_{BN}$ with the consequent shift of the
entropy-bottleneck effect to lower temperature.

\section{Conclusions}

Along this work it was shown how the absence of magnetic order,
inhibiting magnetic degrees of freedom to condensate into a
singlet GS, can be profited to investigate the effects of the
third law of thermodynamics in real materials. It can be observed
how the enhanced paramagnetic correlations strongly increase the
density of low energy excitations inducing the formation of VHF
systems. Apart from the clear quantitative difference of
$C_m/T|_{T\to 0}$ between VHF and NFL, respective power law and
logarithmic thermal dependencies testify their distinct physical
nature.

The divergent increase of this density of excitations collides
with the limited amount of the paramagnetic degrees of freedom
(R$\ln2$ for a doublet GS) producing an entropy-bottleneck which
becomes responsible for the change of the $S_m(T)$ trajectory.
Such 'bottleneck' occurs in these systems because they are not
able to access an ordered state when approaching the exhaustion of
their degrees of freedom. Since this change in the $S_m(T)$
trajectory is not driven by standard magnetic interactions but by
thermodynamic constraints, this process occurs though a continuous
transition of third order character. The temperature at which the
entropy-bottleneck occurs is associated to the $S_m = C_m$
equality.

Notably, among the compounds showing a constant $C_m/T|_{T\to 0}$
also coincident value $\approx 7\pm 0.7$\,J/molK$^2$ is measured.
Despite of the dispersion observed in the $T_{BN}$ values of these
compounds, the convergent value of $C_m/T|_{T\to 0}$ seems to be a
characteristic property which identifies the 'plateau' group
whose. Whether this specific value has an underlying physical
origin remains an open question. Besides that, the coherent regime
observed in $\rho(T)$ around $T_{BM}$ suggests a sort of FL or
eventual SL behavior below that temperature. In standard FL,
constant $C_m/T|_{T\to 0}$ and $\rho(T)$ coherence are related to
a narrow electron band, that in the present scenario could be
tentatively associated to a quantum tunnelling effect connecting
levels of similar energy. This possibility requires to be
confirmed by spectroscopic investigations because it would
represent a singular case of a quantum GS that becomes accessible
due to an entropy bottleneck.

These systems also allow to perform an empirical approach to the
study of the entropy derivatives, showing that negative curvature
($\partial^2S_m/\partial T^2 <0$) has the best physical
representation in NFL compounds, whose logarithmic dependencies of
$S_m(T)$ and $C_m(T)$ place the $S_m=C_m$ equality at $T=0$.
Independently of their NFL or VHF character, the field dependence
of $S_m$ at the milikelvin range is the main parameter that
characterizes the best materials for adiabatic demagnetization
purposes.

It is evident that the recent generation of rare earth based
compounds, exhibiting robust magnetic moments but inhibited to
develop magnetic order, allows to explore the properties of exotic
ground states within a region of temperature where the interplay
between thermodynamic and quantum properties may trigger novel
behaviors.

\section*{Acknowledgments}

The author is grateful to I. Curlik, M. Giovannini, T. Gruner, M.
Deppe, E. Bauer, H. Michor, M. Reiffers, E-W Scheidt, A. Strydom
and I. Zeiringer for allowing to access to original experimental
results. This work was partially supported by projects: PIP-2014
Nr. 112-2013-0100576 of CONICET and SECyT 06/C520 of Univ. of Cuyo
(Arg.).


\begin{thebibliography}{00}
\bibitem{KondoCePdRh} Remind that the case of
CePd$_{0.5}$Rh$_{0.5}$ as an example of QC behavior is included to
show that the change in the $S_m(T)$ trajectory is due to
themodynamic constraints, independently to the cause of the lack
of magnetic order.
\end{thebibliography}
\end{document}